\documentstyle[jkas42]{article}

\runningauthor{S. C. KIM, J. KYEONG, AND E.-C. SUNG}
\runningtitle{OLD OPEN CLUSTER TRUMPLER 5}
\beginpage{100001}
\endpage{100010}

\def\simlt{\lower.5ex\hbox{$\; \buildrel < \over \sim \;$}}
\def\simgt{\lower.5ex\hbox{$\; \buildrel > \over \sim \;$}}
\def\arcdeg{\hbox{$^\circ$}}
\def\arcmin{\hbox{$^\prime$}}
\def\arcsec{\hbox{$^{\prime\prime}$}}
\slugcomment{Not to appear in Nonlearned J., 45. \today}

\begin{document}

\title{NEAR-INFRARED PHOTOMETRIC STUDY OF THE OLD OPEN CLUSTER TRUMPLER 5}
\author{Sang Chul Kim, Jaemann Kyeong, and Eon-Chang Sung}
\offprints{S. C. Kim}
\address{Korea Astronomy and Space Science Institute, Daejeon 305-348, Korea}
\address{\it E-mail: sckim, jman, ecsung@kasi.re.kr}

\vskip 3mm
\address{\normalsize{\it (Received ???. ??, 2009; Accepted ???. ??, 2009)}}

\abstract{
We present $JHK$ near-infrared photometric study
  for the old open cluster (OC) Trumpler 5 (Tr 5), based on the 2MASS data.
From the color-magnitude diagrams of Tr 5, we have located the position 
  of the red giant clump (RGC) stars, and used the mean magnitude 
  of the RGC stars in $K$-band to estimate the distance to Tr 5,
  d $= 3.1 \pm 0.1$ kpc ($(m-M)_0 = 12.46 \pm 0.04$).
From fitting the theoretical isochrones of Padova group,
  we have estimated the reddening, metallicity, and age :
  $E(B-V) = 0.64 \pm 0.05$, [Fe/H] $=-0.4 \pm 0.1$ dex, and
  $t =2.8 \pm 0.2$ Gyr (${\rm log} ~t=9.45 \pm 0.04$), respectively.
These parameters generally agree well with those obtained from the previous studies
  on Tr 5 and confirms that this cluster is an old OC
  with metallicity being metal-poorer than solar abundance,
  located in the anti-Galactic center region.
}


\maketitle
\section{INTRODUCTION}
The recent update of the Galactic open cluster (OC) catalogue by Dias et al. (2002)
  provides 1787 OCs (version 2.10, 2009 February 17),
  while there exist many reports with smaller number of new OCs, such as
  Bica, Dutra, \& Barbuy (2003), 
  Kharchenko et al. (2005) and 
  Froebrich, Scholz, \& Raftery (2007; ``FSR objects'')
  (see also, Lada \& Lada 2003; Kim 2006, and references therein).
Kronberger et al. (2006) discovered 66 stellar groupings 
  from an inspection of the DSS (Digitized Sky Survey) and 2MASS 
  (Two Micron All Sky Survey\footnote{available at
  http://www.ipac.caltech.edu/2mass/releases/ \\ allsky/},
  Skrutskie et al. 1997, 2006) images 
  of selected Milky Way regions,
  whose morphologies, color-magnitude diagrams (CMDs), and stellar density distributions
  suggest that these objects are possible OCs.
Paunzen \& Netopil (2006) have characterized the current status on the accuracy
  of parameters for 395 OCs with more than three independent measurements
  from the literature, and suggested 72 standard OCs with the most accurate
  known parameters which could be served as a standard table in the future
  for testing isochrones and stellar models.

Old OCs, with ages greater than $\sim 1$ Gyr,
  are an important tool for the study of the formation and the evolution 
  of the Galactic disk (Friel 1995; Chen, Hou, \& Wang 2003).
Trumpler 5 (Tr 5; C0634+094, OCL 494, Lund 237, Collinder 105)
  is one of the old OCs
  located in anti-Galactic center region with very small galactic latitude
  (Kim \& Sung 2003, and references therein)
  and among the most massive OCs in the Galaxy (Kaluzny 1998; Cole et al. 2004).
The coordinates of Tr 5 are $\alpha_{J2000}$ = $06^h$ 36$^m$ 42.0$^s$, 
  $\delta_{J2000}$ = $+09\arcdeg~ 25\arcmin~ 58.8\arcsec$,
  $l=$ 202.\arcdeg87, and $b=$ +01.\arcdeg05 (Lyng\r{a} 1987). 
The Trumpler class of Tr 5 is III 1 r, which means 
  (i) Tr 5 is detached and shows no concentration,
  (ii) most stars in this cluster are of nearly the same brightness, and 
  (iii) it is rich.
Table 1 shows the compilation of the previous estimates of
  the fundamental parameters of Trumpler 5.
The studies of Kaluzny (1998), Kim \& Sung (2003), and 
  Piatti, Clari\'a, \& Ahumada (2004) are the only studies 
  based on CCD observations. 
Recently, Cole et al. (2004) and Carrera et al. (2007)
  have used the Ca {\sc{ii}} triplet (8498, 8542, and 8662 \AA) 
  line strengths to measure the metallicity of Tr 5.

Tr 5 is an OC located very close to the Galactic plane ($b=$ +01.\arcdeg05).
Therefore, as Table 1 shows, 
  the mean interstellar reddening toward Tr 5 is $E(B-V)=0.6$,
  which gives optical extinction of $A_V \sim 1.9$ mag.
In the near-infrared (NIR) band, this extinction decreases much to
  $A_J = 0.28 A_V = 0.5$ mag and $A_K = 0.11 A_V = 0.2$ mag
  (Cardelli, Clayton, \& Mathis 1989).
Tr 5 is also known to have differential reddenings (Piatti et al. 2004).
NIR photometric study, therefore, is a better way to study this cluster than 
  the studies using optical photometry data.
In this paper, we have analyzed the NIR photometry data of Tr 5
  obtained from the 2MASS project.

Section II describes the NIR data sets used in this study and
Sections III presents the CMDs of Tr 5.
Sections IV, V, and VI present the distance, Padova isochrone fitting results, 
  and the color-color diagram, respectively, for the OC Tr 5.
Finally, a summary is given in Section VII.

\begin{table*}[t]
\begin{center}
{\bf Table 1.}~~A List of the Estimation of the Physical Parameters of Trumpler 5 \\
\vskip 3mm
{\small
\setlength{\tabcolsep}{1.2mm}
\begin{tabular}{lll} \hline\hline
Parameter & Information & Reference \\
\hline 
Reddening, $E(B-V)$ & $0.80$ mag    & Dow \& Hawarden 1970; Janes \& Adler 1982 \\
                    & $0.48$ mag    & Kalinowski 1974; Kalinowski et al. 1974 \\
                    & $0.60$ mag    & Piccirillo et al. 1977 \\
                    & $0.64$ mag    & Kalinowski 1979; Janes \& Adler 1982 \\
                    & $0.58$ mag    & Kaluzny 1998 \\
                    & $0.60 \pm 0.10$ mag    & Kim \& Sung 2003 \\
                    & $0.66         $ mag           & Cole et al. 2004 \\
                    & $0.60 \pm 0.04$ mag    & Piatti et al. 2004 \\
                    & $0.64 \pm 0.05$ mag    & This study \\
mean                & $0.62 \pm 0.08$ mag    & \\
\hline 
Distance, d  & $2.4        $ kpc ($(m-M)_0$ = $11.92 $ mag) & Dow \& Hawarden 1970; 
                                                             Janes \& Adler 1982 \\
             & $2.9 \pm 0.1$ kpc ($(m-M)_0$ = $12.3  \pm 0.1 $ mag)  & Kalinowski 1974;
                                                              Kalinowski et al. 1974 \\
             & $2.4 \pm 0.3$ kpc ($(m-M)_0$ = $11.9  \pm 0.3 $ mag) & Kalinowski 1975 \\
             & $1.9        $ kpc ($(m-M)_0$ = $11.4          $ mag) & Piccirillo et al. 1977\\
             & $1.0        $ kpc ($(m-M)_0$ = $10.02         $ mag)  & Kalinowski 1979;
                                                             Janes \& Adler 1982 \\
             & $3.0        $ kpc ($(m-M)_0$ = $12.4          $ mag)  & Kaluzny 1998 \\
             & $3.4 \pm 0.3$ kpc ($(m-M)_0$ = $12.64 \pm 0.20$ mag)  & Kim \& Sung 2003 \\
             & $2.8        $ kpc            & Cole et al. 2004 \\
             & $2.4 \pm 0.5$ kpc ($(m-M)_0$ = $11.88 \pm 0.30$ mag) & Piatti et al. 2004 \\
             & $3.1 \pm 0.1$ kpc ($(m-M)_0$ = $12.46 \pm 0.04$ mag) & This study \\
mean         & $2.5 \pm 0.7$ kpc ($(m-M)_0$ = $11.9  \pm 0.8 $ mag) & \\
\hline 
Age  & $4.9$ Gyr           (from the morphological age index) &
                  Janes \& Phelps 1994; Cole et al. 2004 \\
     & $4.1        $ Gyr & Kaluzny 1998 \\
     & $2.4 \pm 0.2$ Gyr (${\rm log} ~t=9.38 \pm 0.04$) & Kim \& Sung 2003 \\
     & $5.67\pm2.26$ Gyr & Salaris et al. 2004 \\
     & $3.0 \pm 0.5$ Gyr           & Cole et al. 2004 \\
     & $5.0 \pm 0.5$ Gyr & Piatti et al. 2004 \\
     & $2.8 \pm 0.2$ Gyr (${\rm log} ~t=9.45 \pm 0.04$) & This study \\
mean & $4.0 \pm 1.2$ Gyr (${\rm log} ~t=9.60 \pm 0.15$) & \\
\hline 
Metallicity, [Fe/H] & $ 0.0          $ dex & Kaluzny 1998 \\
                    & $-0.30 \pm 0.10$ dex & Kim \& Sung 2003 \\
                    & $-0.56 \pm 0.11$ dex$^\dagger$ & Cole et al. 2004 \\
                    & $-0.30 \pm 0.15$ dex & Piatti et al. 2004 \\
                    & $-0.36 \pm 0.05$ dex$^\ddagger$ & Carrera et al. 2007 \\
                    & $-0.4  \pm 0.1 $ dex            & This study \\
mean                & $-0.32 \pm 0.17$ dex           & \\
\hline
\end{tabular}
} 
\end{center} 
\hspace{0.0cm}$^{\dagger}$ : From Ca {\sc{ii}} triplet lines of 10 RGB member stars \\
\hspace{-1.0cm}$^{\ddagger}$ : From Ca {\sc{ii}} triplet lines of 21 member stars \\
\end{table*}

\section{THE 2MASS DATA}
The 2MASS project (Skrutskie et al. 2006) have used
  two dedicated 1.3 m telescopes
  located at Mount Hopkins, Arizona, and Cerro Tololo, Chile and
  $256 \times 256$ NICMOS3 (HgCdTe) arrays manufactured
  by Rockwell International Science Center (now Rockwell Scientific),
  which give field-of-view of $8.'5 \times 8.'5$ and pixel scale of
  $2\arcsec$ pixel$^{-1}$ (Kim 2006).
The photometric system comprise $J$ (1.25 $\mu$m),
  $H$ (1.65 $\mu$m) and $K_S$ (2.16 $\mu$m) bands, where
  the ``$K$-short'' ($K_S$) filter excludes wavelengths longward of 2.31 $\mu$m
  to reduce thermal background and airglow and includes wavelengths as short as
  2.00 $\mu$m to maximize bandwidth (see Figure 2 of Skrutskie et al. 2006 or
  Figure 7 of Bonatto, Bica, \& Girardi 2004
  for the transmission curves of the 2MASS filters; Carpenter 2001).

VizieR\footnote{http://vizier.u-strasbg.fr/viz-bin/VizieR?-source=2MASS}
  was used to extract $J, H,$ and $K_S$ 2MASS photometry data
  in circular areas centered on the cluster Tr 5.
Figure 1 displays the grey-scale image of Tr 5
  taken from the DSS.
2MASS $J$, $H$, and $K_S$-band images covering the field of Tr 5
  are shown in Figures 2, 3, and 4, respectively.

The distribution of photometric errors is shown in Figure 5.
2MASS photometric errors typically attain 0.1 mag at $J \approx 16.0$ mag,
  $H \approx 15.4$ mag, and $K_S \approx 14.6$ mag.
These photometric uncertainties are of high enough quality, 
  comparable to other NIR photometric studies (see, e.g., 
  Bonatto, Bica, \& Santos (2005), Bonatto \& Bica (2007);
  see also Sung, Stauffer, \& Bessell (2009)
  for the distribution of photometric errors of 
  the {\it Spitzer Space Telescope} Infrared Array Camera (IRAC) 
  undersampled data).

\begin{figure}[p]
  \epsfxsize=7.5cm
  \centerline{\epsffile{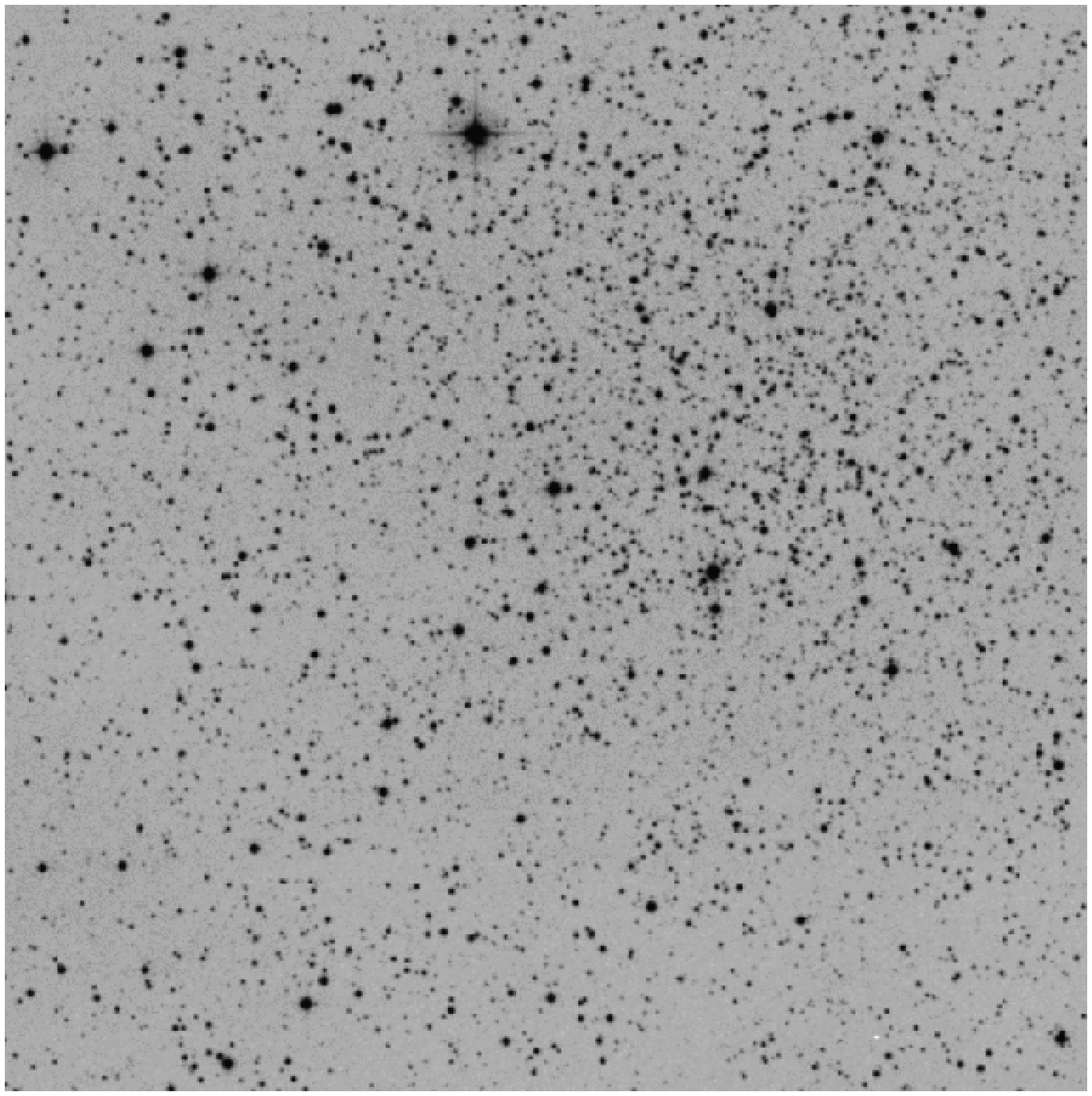}}
{\small {\bf ~~~Fig. 1.}---~Second generation DSS optical 
  red image of the cluster
  Tr 5.  The field of view is $15\arcmin \times 15\arcmin$.
North is up and east is to the left.
The coordinate of the cluster center is
  ($\alpha_{J2000}$, $\delta_{J2000}$) =
  ($06^h$ 36$^m$ 42.0$^s$, $+09\arcdeg~ 25\arcmin~ 58.8\arcsec$).
}
\end{figure}

\begin{figure}[p]
  \epsfxsize=7.5cm
  \centerline{\epsffile{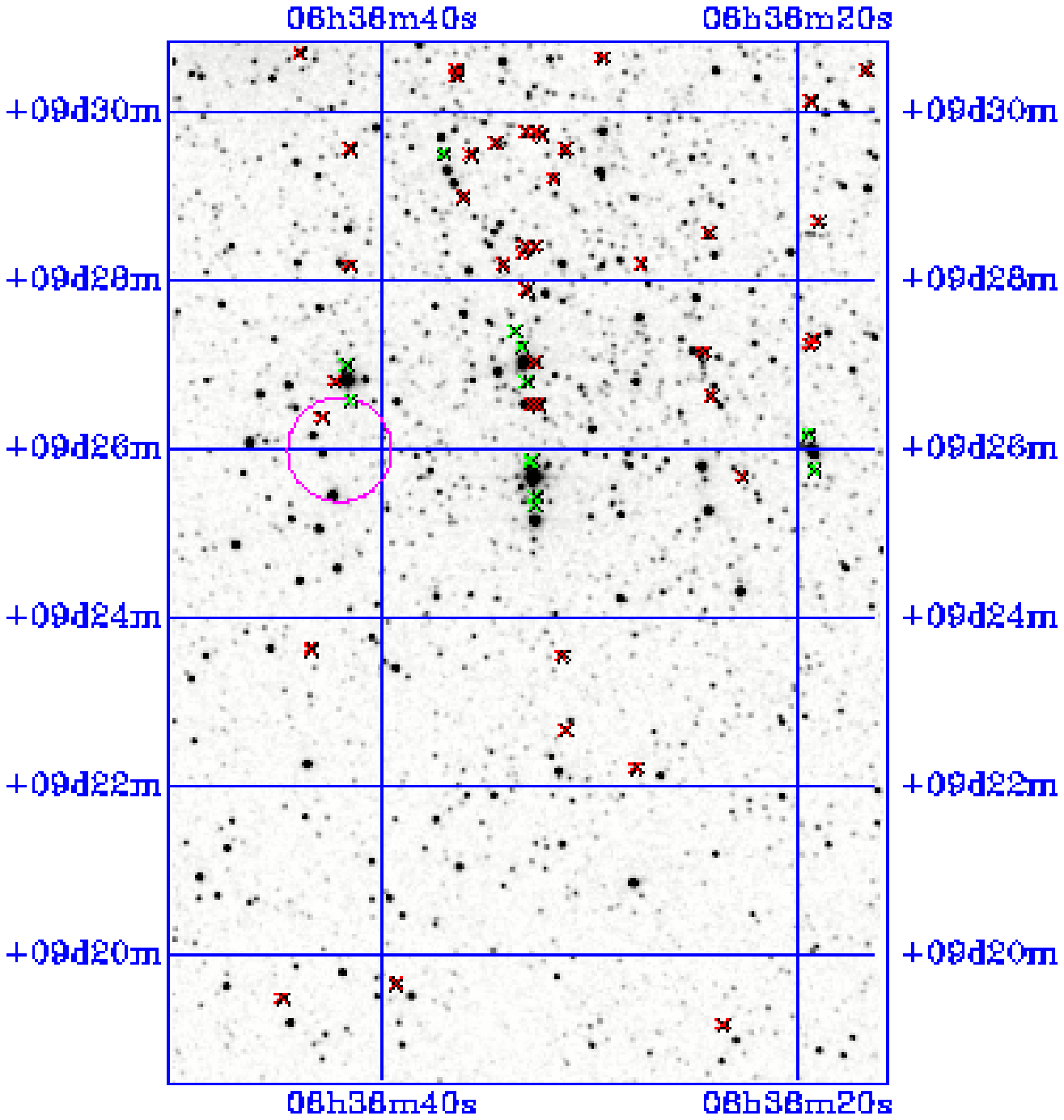}}
{\small {\bf ~~~Fig. 2.}---~2MASS $J$-band image of the cluster Tr 5,
  provided by the 2MASS Image Service.
The open circle indicates the object central part.
Crosses indicate instrumental artifacts.
}
\end{figure}

\begin{figure}[p]
  \epsfxsize=7.5cm
  \centerline{\epsffile{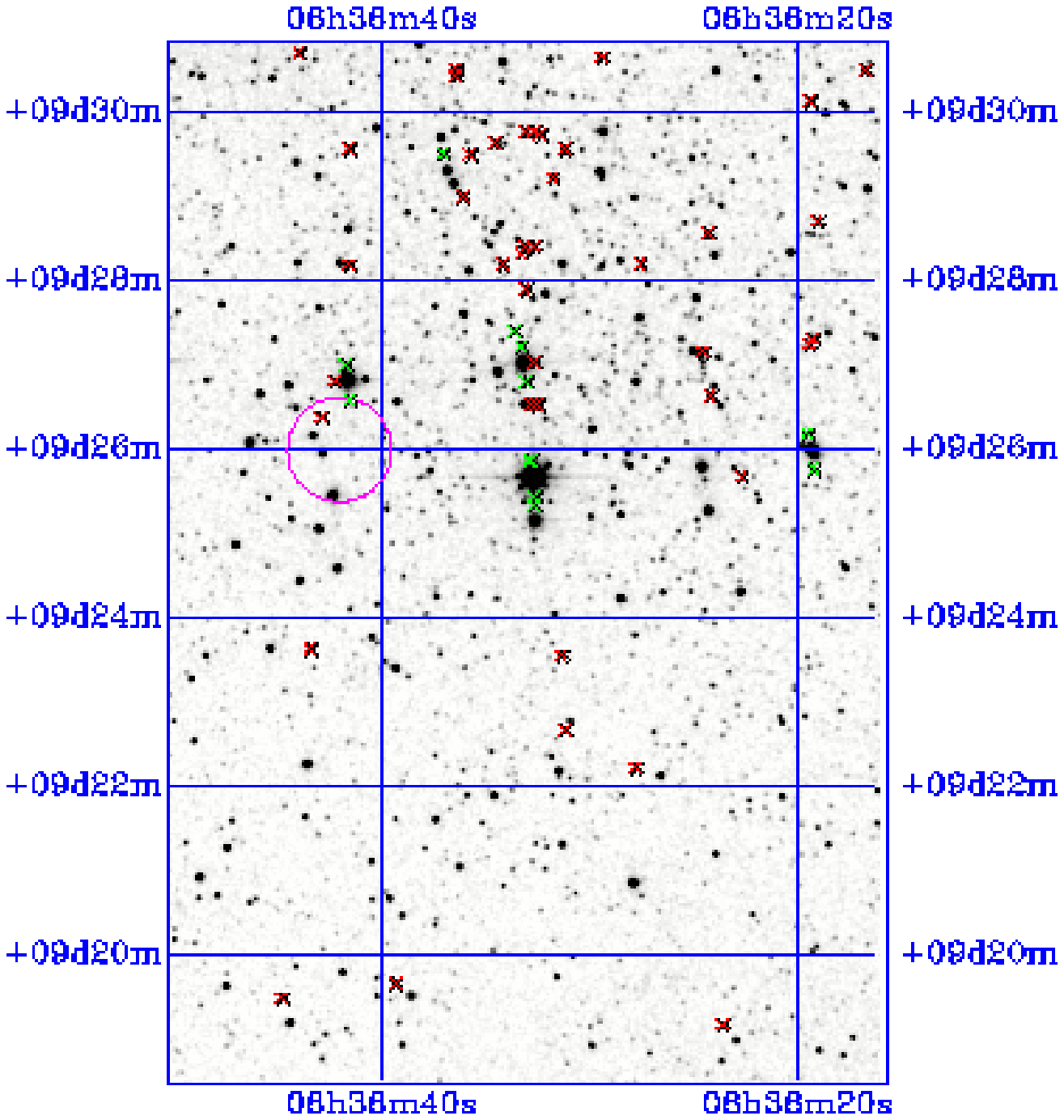}}
{\small {\bf ~~~Fig. 3.}---~2MASS $H$-band image of the cluster Tr 5,
  provided by the 2MASS Image Service.
The open circle indicates the object central part.
Crosses indicate instrumental artifacts.
}
\end{figure}

\begin{figure}[p]
  \epsfxsize=7.5cm
  \centerline{\epsffile{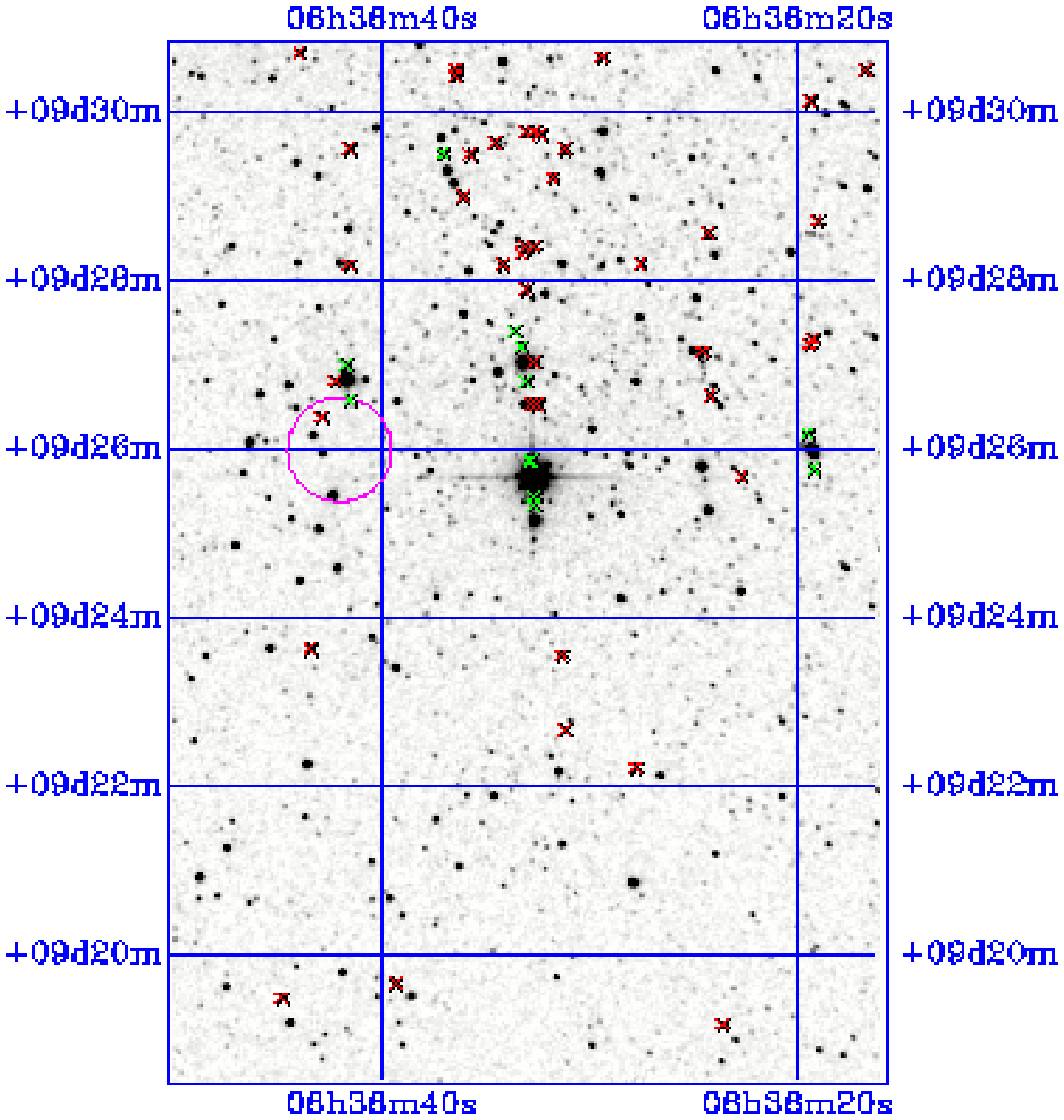}}
{\small {\bf ~~~Fig. 4.}---~2MASS $K_S$-band image of the cluster Tr 5,
  provided by the 2MASS Image Service.
The open circle indicates the object central part.
Crosses indicate instrumental artifacts.
}
\end{figure}

\begin{figure}[p]
  \epsfxsize=8.5cm
  \centerline{\epsffile{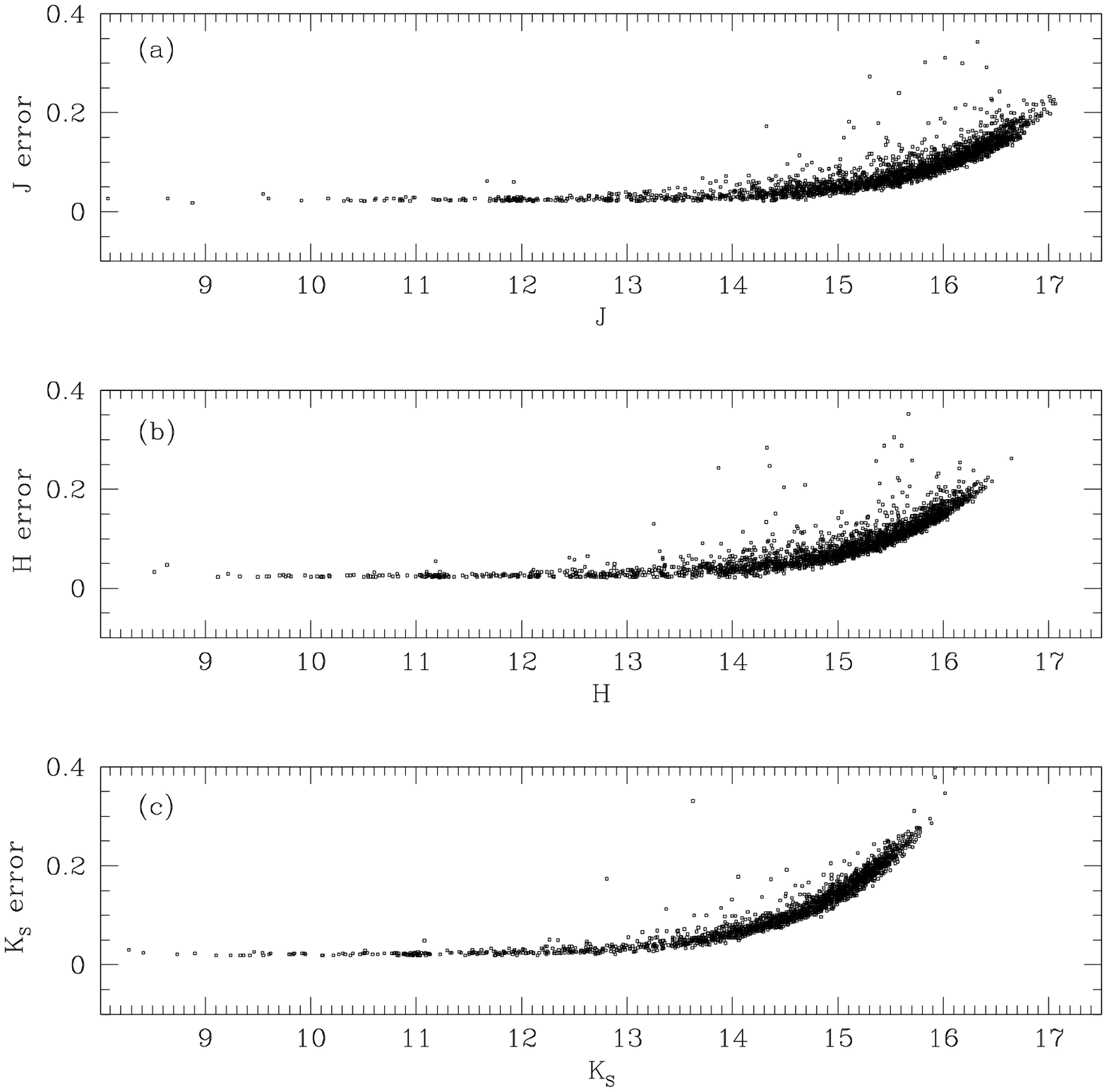}}
{\small {\bf ~~~Fig. 5.}---~Distribution of the $J$, $H$, and 
  $K_S$-band photometric errors as a function of magnitude.
}
\end{figure}

\section{THE COLOR-MAGNITUDE DIAGRAMS}
Figure 6 shows the $J$ vs. $(J-H)$ (panel a) and 
  $K_S$ vs $(J-K_S)$ (panel b) CMDs for the Tr 5 stars 
  in radial region of R $<7.7\arcmin$.
From the stellar density profile of Tr 5 stars observed with the Cerro Tololo
  Inter-American Observatory (CTIO) 0.9 m telescope,
  Piatti et al. (2004) have measured the angular radius of Tr 5 
  as R $\sim 7.7\arcmin \pm 0.3\arcmin$, which is used to plot Figure 6.
Thick band of many stars are seen below the main-sequence turn-off (MSTO).
Broad red giant branch (RGB) and red giant clump 
  (RGC; $J_{RGC} = 11.9 \pm 0.1$, $K_{S,RGC} = 11.0 \pm 0.1$,
  $(J-H)_{RGC} = 0.73 \pm 0.03$, and $(J-K_S)_{RGC} = 0.90 \pm 0.05$) 
  can also be found.

\begin{figure}[p]
  \epsfxsize=8.5cm
  \centerline{\epsffile{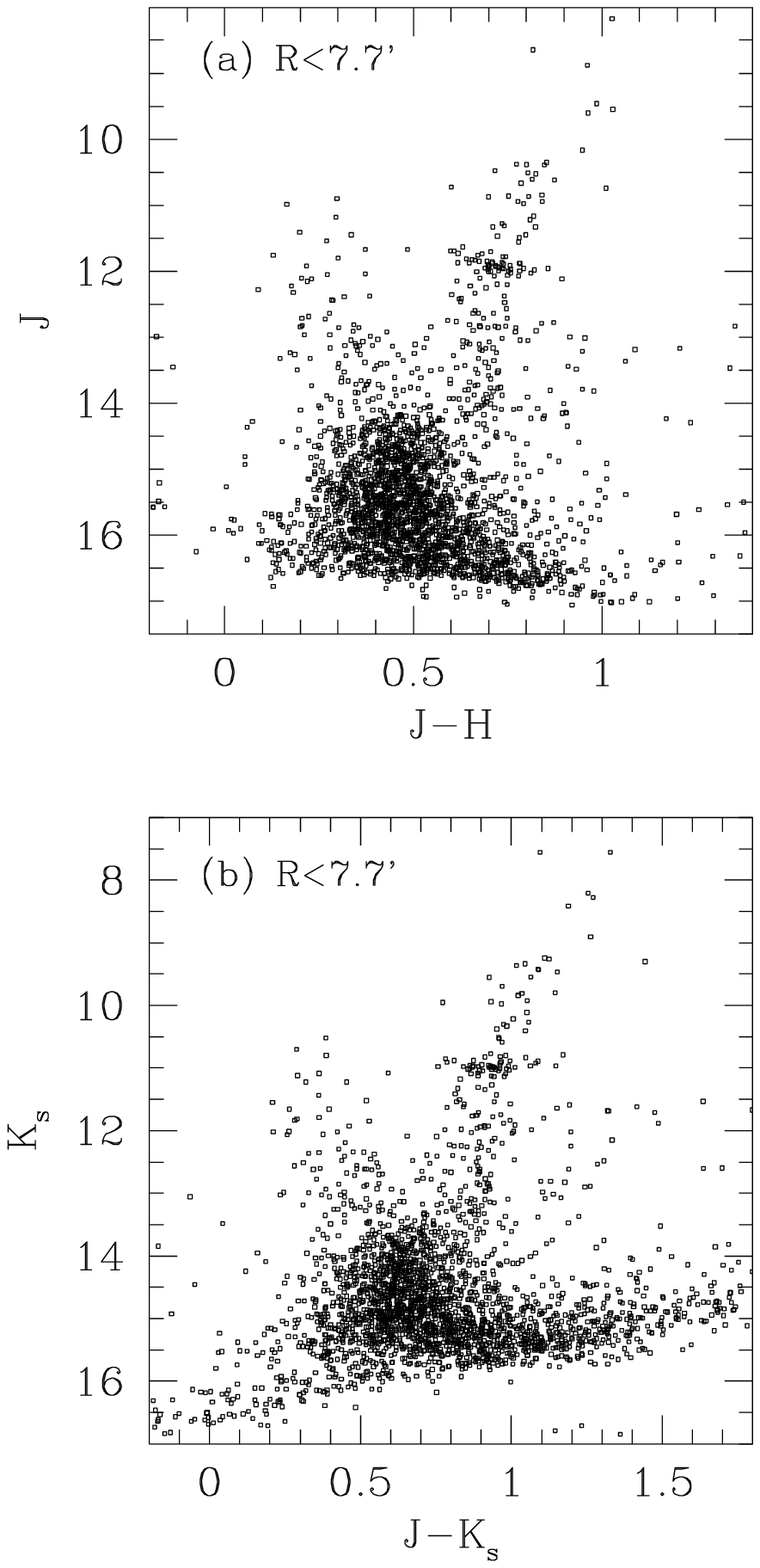}}
{\small {\bf ~~~Fig. 6.}---~(a) $J$ vs. $(J-H)$ and
  (b) $K_S$ vs $(J-K_S)$ CMDs for the stars
  in the field (R $<7.7\arcmin$) of Tr 5.
}
\end{figure}

Several stars are located at the blue straggler (BS) region above the MSTO
  and bluer than $J-H \approx 0.5$ and $J-K_S \approx 0.65$.
Since Tr 5 is located at the Galactic plane ($b=$ +01.\arcdeg05),
  it is needed to get detailed membership studies for Tr 5
  based either on radial velocity or proper motion
  to make a robust BS list with as least as possible field star contamination
  (Ahumada \& Lapasset 1995; De Marchi et al. 2006; Ahumada \& Lapasset 2007; 
  Carraro, V\'azquez, \& Moitinho 2008; Kyeong et al. 2008; Momany et al. 2008).
We have statistically tested the possible number of BS stars in the Tr 5 field
  by comparing the CMDs of Tr 5 and two nearby control fields.
Two control fields are chosen at the same Galactic latitude,
  but with one degree larger and one degree smaller Galactic longitude
  for control field 1 and control field 2, respectively, 
  than that of Tr 5.
From the comparison of the CMDs of Tr 5 and the control fields
  for a given magnitude and color range,
  we counted the number of stars in a control field and
  subtracted this number of stars in the Tr 5 CMD randomly.
  
In Figure 7, we showed the raw CMD of R$<6\arcmin$ field of Tr 5 (panel (a)),
  two CMDs of the control fields 1 and 2 (panels (b) and (c), respectively), and
  the field star eliminated CMDs using the control fields 1 and 2 
  (panels (d) and (e), respectively).
The solid boxes in Figure 7 are plotted to estimate the number of stars
  in the BS region.
There are 29 stars in the box of Fig. 7 (a), 
  20 and 23 stars in those of Fig. 7 (b) and (c), respectively, and
  14 and 12 stars remain in those of Fig. 7 (d) and (e), respectively, after 
  the elimination of the field star contamination.
This indicates that $\simgt 40$ \% stars in the BS region of Tr 5 CMD
  possibly could be bona-fide BSs,
  although the number of stars in the BS region increases at outer regions
  than the center (Figures 8 and 9).

\begin{figure}[p]
  \epsfxsize=8.5cm
  \centerline{\epsffile{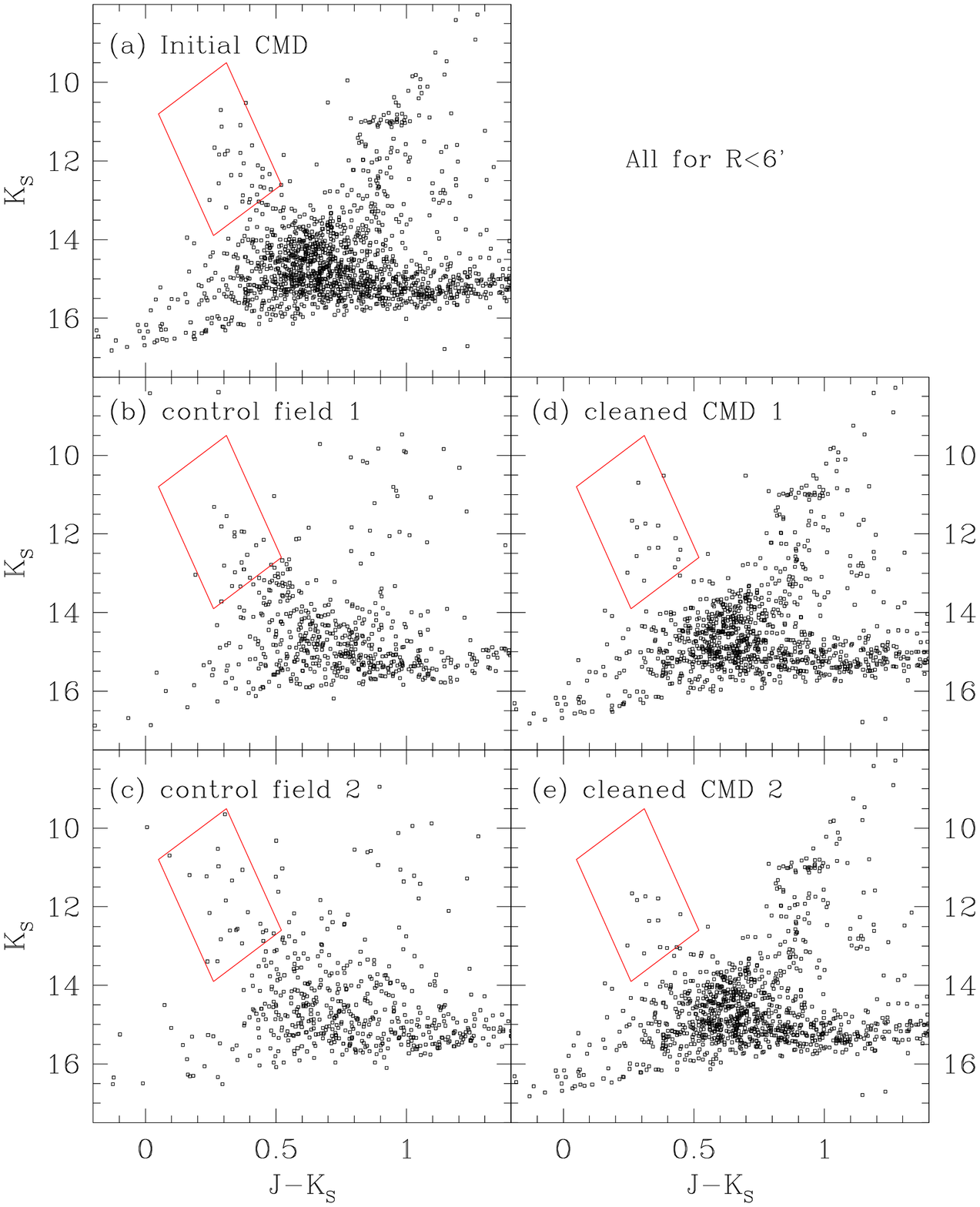}}
{\small {\bf ~~~Fig. 7.}---~Statistical elimination of field star
  contamination.
(a) $K_S$ vs. $(J-K_S)$ CMDs for the stars in the R$<6\arcmin$ field 
  of Tr 5.
(b) and (c) are CMDs for the control fields 1 and 2 (all for R$<6\arcmin$), 
  respectively,
  and (d) and (e) are field star contamination eliminated CMDs
  using the control fields 1 and 2, respectively.
The solid boxes are plotted to estimate the number of stars
  in the BS region.
}
\end{figure}

From the wide width of the MS and the tilted placement of the RGC (following
  the reddening vector) in the CMDs of Tr 5,
  Piatti et al. (2004) have shown that Tr 5 has non-uniform extinctions
  [$\Delta E(B-V) = 0.11 - 0.22$] over the face of the cluster.
Piatti et al. (2004) showed that 
  the stars with relatively small photometric errors and/or the lowest color excesses
  are mostly distributed in the cluster core region (R $< 2\arcmin$) and 
  in the northeastern half of their observed field.
They also showed that the highest reddened group of stars is found towards the south
  and there is an obscure cloud of interstellar matter as large as 
  a quarter of ring of $\sim 2\arcmin$ wide 
  at $3\arcmin$ from the cluster center towards the southwestern side.
Figure 6 (a) clearly reveals that the RGC is tilted and extended
  and the RGB has wide width.
These facts reveal the existence of the differential reddening 
  in the field of Tr 5,
  while it is not so clear in the CMD of $K_S$ vs $(J-K_S)$ (Figure 6 (b))
  which passband is less affected by the reddenings.

\section{DISTANCE}  
The helium-burning RGC stars are known to have great potential to be used
  as a standard candle and
  there are claims that their absolute magnitude in a certain band is constant
  or that the luminosity of RGC is a function of age and/or metal abundance
  (see, e.g., Grocholski \& Sarajedini 2002).
Janes \& Phelps (1994) estimated the mean color and magnitude
  of the RGC in old open clusters to be
  $(B-V)_{0, RGC} = 0.95 \pm 0.10$, and $M_{V, RGC} = 0.90 \pm 0.40$,
  when the $V$ magnitude difference between the RGC and
  the MSTO of the clusters, $\delta V$, is greater than one.
Using {\it Hipparcos} RGC stars with parallax errors of less than 10\%
  to calculate the $I$-band absolute magnitude of the solar neighborhood RGC,
  Paczy\'nski \& Stanek (1998) find $M_{I,RGC} = -0.28 \pm 0.09$ mag 
  having no variation with color over the range $0.8 < (V-I)_0 <1.4$,
  and Stanek \& Garnavich (1998) find a similar result with 
  $M_{I,RGC} = -0.23 \pm 0.03$ mag.
From the $K$-band luminosity of 238 RGC stars,
  Alves (2000) finds $M_{K,RGC} = -1.61 \pm 0.03$ mag
  with no correlation between [Fe/H] and $M_K$, and
  Grocholski \& Sarajedini (2002) find $M_{K,RGC} = -1.62 \pm 0.06$ mag
  from the $JK$ photometry of 14 OCs.
The other advantage that the $K$-band holds is the relatively small 
  amount of reddening effect in this band, therefore making it
  relatively free of the differential reddening effect
  (seen noticeably by the tilt of the RGC stars 
  in the optical CMDs in Kim \& Sung (2003) and $J$ vs. $(J-H)$ CMDs in Figure 8)
  and efficient in determining the mean magnitude of the RGC stars.
  
If we convert the 2MASS $K_S$-band to the $K$-band 
  adopting the Bessell \& Brett (1988) system 
  using the transformation equation A1 of Carpenter (2001),
  \begin{equation}
  K_{BB} = K_S - (-0.044\pm0.003) -(0.000\pm0.005)(J-K_{BB}),
  \end{equation}
  we get the magnitude of the RGC obtained in the previous section to be
  $K_{RGC} = 11.04 \pm 0.06$ mag.  
If we adopt the absolute magnitude of the RGC stars of Grocholski \& Sarajedini (2002),
  $M_{K,RGC} = -1.62 \pm 0.06$ mag,
  we get the apparent distance modulus of Tr 5, $(K-M_K)_{RGC} = 12.66  \pm 0.04 $ mag
  and the true distance modulus of
  $(m-M)_0 = (K-M_K)_{RGC} - A_K = 12.46  \pm 0.04 $ mag
  (d $= 3.1 \pm 0.1$ kpc), using $A_K = 0.11 A_V$.

\section{PADOVA ISOCHRONE FITTING}  
In Figure 8, we have plotted the $J$ vs. $(J-H)$ CMDs for the Tr 5 stars
  in different radial regions
  as indicated in each panel from (a) R $< 2\arcmin$ to (f) R $<7.7\arcmin$.
We have fitted the theoretical Padova isochrones 
  computed with the 2MASS $J, H$ and $K_S$ filters (Bonatto, Bica, \& Girardi 2004;
  Bica, Bonatto, \& Blumberg 2006)
  to derive the cluster parameters (Kim 2006).
As Table 1 shows that the $E(B-V)$ reddening values toward Tr 5
  obtained from the previous studies (especially the recent studies) 
  converge to $\sim 0.6$ mag,
  we also obtained the best fit using this reddening value of 
  $E(B-V) = 0.60 \pm 0.10$ mag in Figure 8
  (using $E(J-H)=0.309 E(B-V)$, $E(J-K_S)=0.488 E(B-V)$).
The CMDs for the stars in the central regions (panels (a) and (b))
  show best fits with the Padova isochrones.
The CMDs, however, for stars in the intermediate regions (panels (c) and (d)) 
  start to show hint of RGC tilt and RGB broadening, and
  those for stars in the outer regions (panels (e) and (f))
  clearly show the effect of differential reddening (Piatti et al. 2004).
The Padova isochrone fitting using the reddening value of $E(B-V) = 0.60 \pm 0.10$ mag
  and the distance modulus
  obtained in the previous section ($(m-M)_0 = 12.46 \pm 0.04$ mag)
  gives the best match with 
  metallicity [Fe/H] $= -0.4 \pm 0.1$ dex and 
  age$= 2.8 \pm 0.2$ Gyr (${\rm log} ~t=9.45 \pm 0.04$).
Most of the recent studies after the year of 2000 give the age for Tr 5
  between $\sim 3$ and 5 Gyr and
  the metallicity for Tr 5 between [Fe/H] $\sim -0.30$ and $-0.50$ dex.
This confirms that Tr 5 is one of the oldest 
  OCs in the Galaxy (see, e.g., Bonatto et al. 2006; Bonatto \& Bica 2007;
  Froebrich, Meusinger, \& Davis 2008; Momany et al. 2008; Froebrich et al. 2009),
  and that this cluster is a metal-poor OC
  not being a solar-metallicity cluster (Kaluzny 1998; Piatti et al. 2004).

In Figure 9, we have showed the Padova isochrone fittings for the
  $K_S$ vs $(J-K_S)$ CMDs, as in Figure 8,
  in different radial regions
  as indicated in each panel from (a) R $< 2\arcmin$ to (f) R $<7.7\arcmin$.
Although all the same parameters are used for other parameters,
  somewhat larger reddening value ($E(B-V)=0.67 \pm 0.02$) was used,
  because a poor fit is resulted from the reddening value of $E(B-V) = 0.60$.
This slight discrepancy seen in best-fit parameters among different diagrams
  might be caused by the uncertainties in the photometry and/or isochrones
  (Rider et al. 2004).
We used the mean of the reddening values obtained from the Padova isochrone fittings for
  the $J$ vs. $(J-H)$ and $K_S$ vs $(J-K_S)$ CMDs,
  $<E(B-V)> = 0.64 \pm 0.05$ in Table 1.

\begin{figure*}[p]
  \epsfxsize=16cm
  \centerline{\epsffile{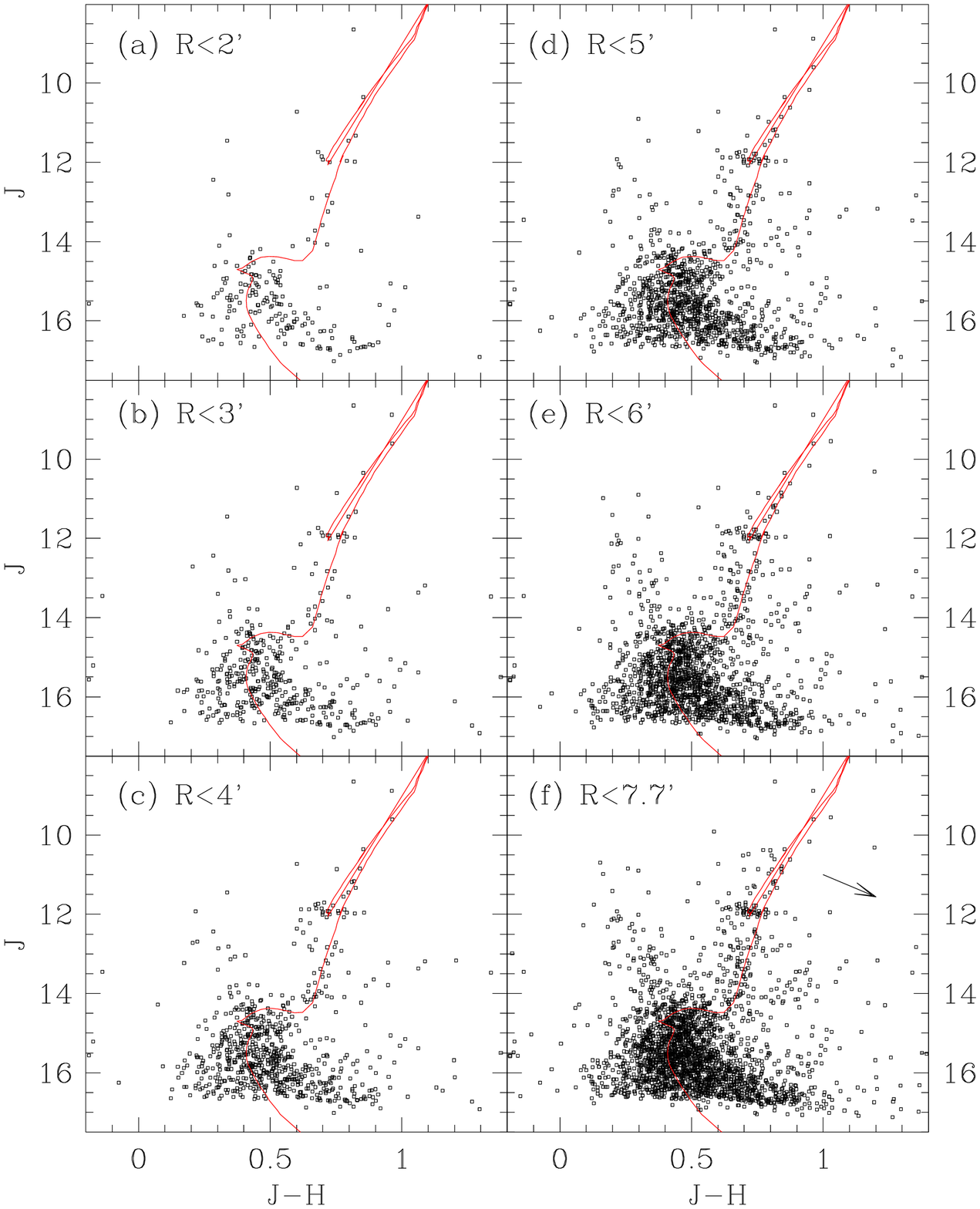}}
{\small {\bf ~~~Fig. 8.}---~$J$ vs. $(J-H)$ CMDs for the stars
  in the field of Tr 5.
The solid lines represent the Padova isochrones (Bonatto, Bica, \& Girardi 2004)
  for $E(B-V)=0.60$, $J-M_J = 12.98$,
  Z$=0.008$, and age = 2.8 Gyr.
The reddening vector with $E(J-H)=0.20$ is shown in panel (f),
  where $A_J = 0.28 A_V$, $R_V = 3.1$ and $E(J-H)=0.309 E(B-V)$ are used.
}
\end{figure*}

\begin{figure*}[p]
  \epsfxsize=16cm
  \centerline{\epsffile{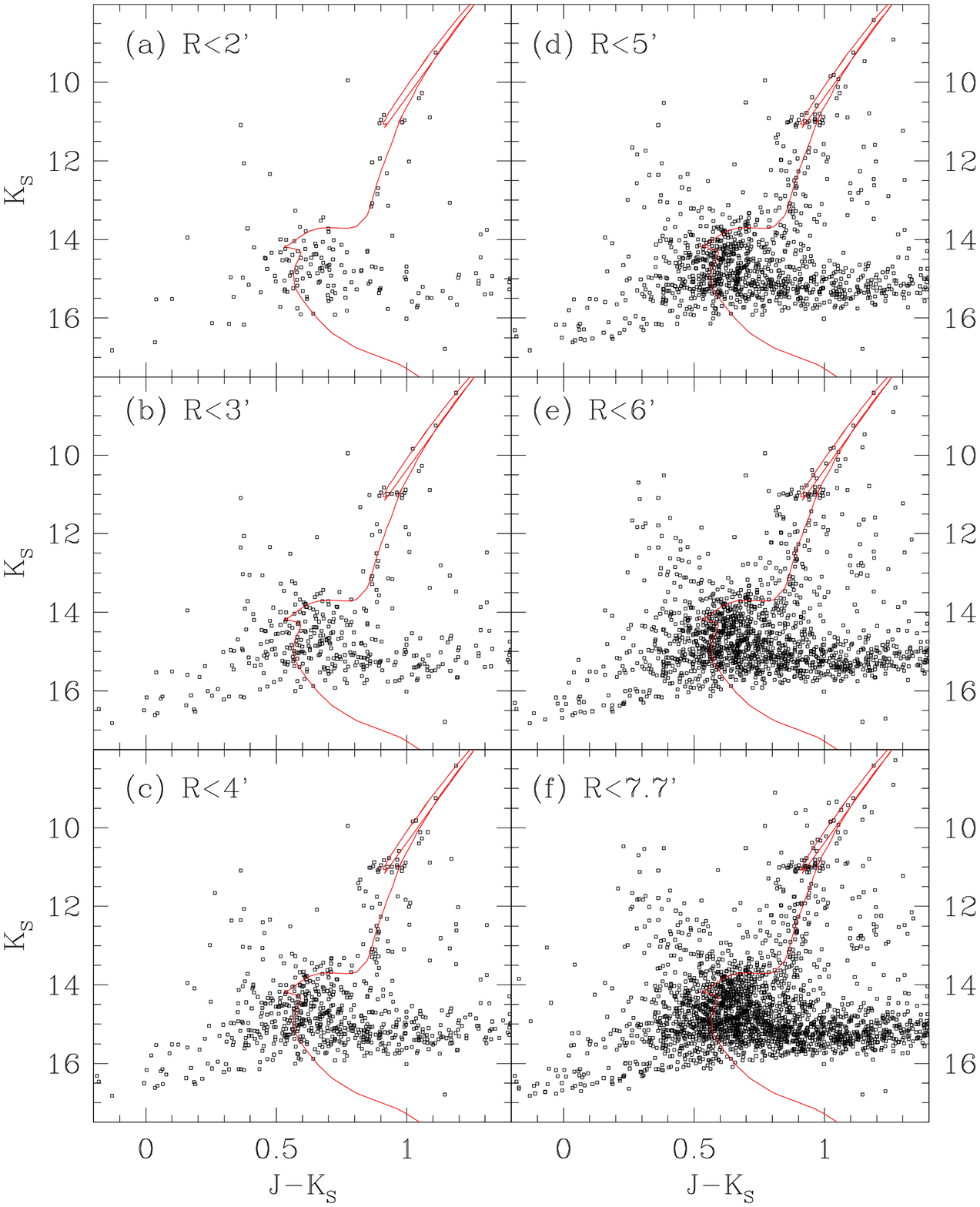}}
{\small {\bf ~~~Fig. 9.}---~$K_S$ vs. $(J-K_S)$ CMDs for the stars
  in the field of Tr 5.
The solid lines represent the Padova isochrones (Bonatto, Bica, \& Girardi 2004)
  for $E(B-V)=0.67$, $K_S-M_{K_S} = 12.66$,
  Z$=0.008$, and age = 2.8 Gyr.
}
\end{figure*}

\section{COLOR-COLOR DIAGRAM} 
Figure 10 shows the $(J-H) \times (H-K_S)$ color-color diagram of
  the stars with $K_S$ magnitude errors less than 0.1 mag in Tr 5
  (filled circles).
The MS range of the reddened Padova isochrones are overplotted
  as solid line ($0.15 - 1.4 M_\odot$) and
  a reddening vector with $E(J-H) = 1.72 \times E(H-K_S)$
  for $A_J = 0.55$ is denoted as an arrow.
The locus of unreddened MS stars in the 2MASS system
  taken from Sung et al. (2008) is denoted by long-dashed line.
The loci of unreddened MS and giants stars
  in the Bessell \& Brett (1988) system
  (not converted to the 2MASS system) taken from Bessell \& Brett (1988)
  are denoted by dotted and short-dashed lines, respectively.

Most of the stars in Tr 5 plotted in Figure 10 are distributed along
  the main sequences of the Padova isochrone, but
  in the locations affected by the reddenings
  of the amount of $A_V = 1.98$ and $A_J = 0.55$.
The difference between the locus of the unreddened MS stars 
  of Sung et al. (2008) and that of the reddened Padova isochrones 
  is in very good agreement with the reddening value for Tr 5
  denoted by the arrow.
While Sung et al. (2008) have derived the mean locus of MS stars
  in the 2MASS system from compilation of 2MASS data 
  for the late type MS stars in the solar neighborhood and 
  various OCs,
  it would be highly anticipated to derive that for giant stars
  in the 2MASS system.

\begin{figure}[p]
  \epsfxsize=8.5cm
  \epsfysize=8.5cm
  \centerline{\epsffile{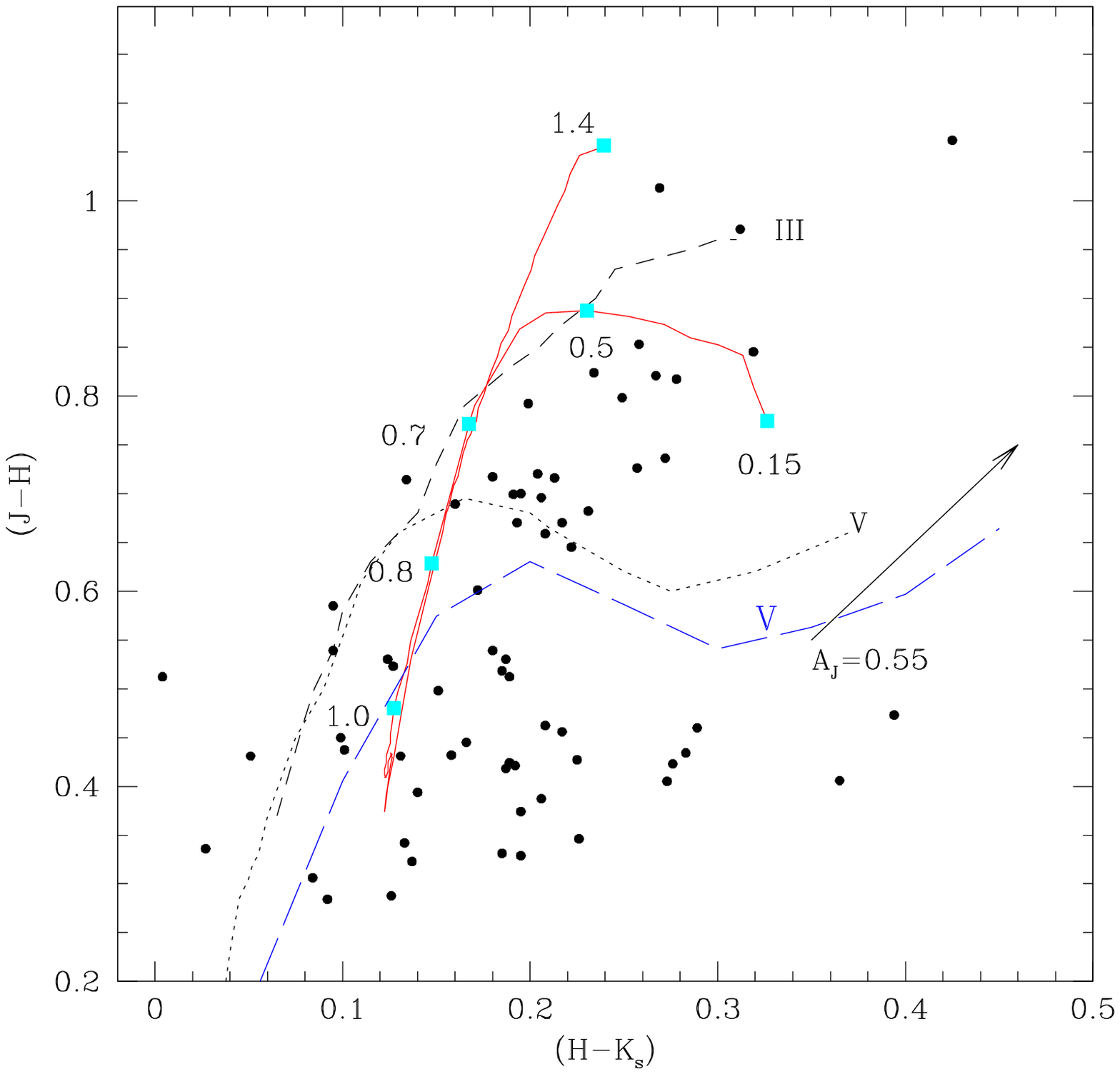}}
{\small {\bf ~~~Fig. 10.}---~$(J-H) \times (H-K_S)$ color-color diagram 
  of the stars in Tr 5 for $K_S$ magnitude errors less than 0.1 mag
  (filled circles).
The MS range of the reddened Padova isochrones are overplotted.
Solid line is the MS range of the Padova isochrones of age 2.8 Gyr and
  representative masses in $M_\odot$ are indicated along the isochrone.
Long-dashed line is the locus of unreddened MS stars in the 2MASS system
  taken from Sung et al. (2008).
Dotted and short-dashed lines are the loci of unreddened MS and giants
  stars, respectively, in the Bessell \& Brett (1988) system,
  not converted to the 2MASS system, taken from Bessell \& Brett (1988).
Arrow is a reddening vector with $E(J-H) = 1.72 \times E(H-K_S)$ for
  $A_V = 1.98$ and $A_J = 0.55$.
}
\end{figure}

\section{SUMMARY} 
We presented the analysis of the $JHK_S$ near-infrared photometry
  for the old OC Tr 5 using the 2MASS data.
We used these 2MASS $JHK_S$ photometry data for the study of Tr 5
  since this cluster is located very close to the Galactic plane 
  ($b=$ +01.\arcdeg05) and has high ($E(B-V) \sim 0.6$) 
  interstellar reddening values.
Since the $K$-band is relatively free of the differential reddening
  effect which exists in the face of Tr 5,
  the distance to Tr 5 is easily obtained using the $K$-band
  magnitude of the RGC of this cluster.
The CMDs plotted for different radial regions clearly showed
  the differential reddening over the face of Tr 5 (Piatti et al. 2004).
The primary results obtained from the $J$ vs. $(J-H)$ and $K_S$ vs. $(J-K_S)$ 
  CMDs of Tr 5 are as follows :

1. The RGC stars are located at 
  $J_{RGC} = 11.9 \pm 0.1$, $K_{S,RGC} = 11.0 \pm 0.1$,
  $(J-H)_{RGC} = 0.73 \pm 0.03$, and $(J-K_S)_{RGC} = 0.90 \pm 0.05$ and
  the distance to Tr 5 is obtained (d $= 3.1 \pm 0.1$ kpc,
  $(m-M)_0 = 12.46 \pm 0.04$) using the mean magnitude of the RGC stars 
  in $K_S$-band.

2. From the Padova isochrone fittings to the CMDs,
  we have estimated the reddening, metallicity, and age :
  $E(B-V) = 0.64 \pm 0.05$, [Fe/H] $=-0.4 \pm 0.1$ dex, and
  age $=2.8 \pm 0.2$ Gyr (${\rm log} ~t=9.45 \pm 0.04$), respectively.
These parameters confirm that Tr 5 is an old OC with
  metallicity being metal-poorer than solar abundance,
  located in the anti-Galactic center region.

\vspace{4mm}
The authors are grateful to the anonymous referee for numerous comments and
  suggestions that improved the quality of this manuscript.
This publication makes use of data products from the Two Micron All Sky Survey,
  which is a joint project of the University of Massachusetts and
  the Infrared Processing and Analysis Center/California Institute of Technology,
  funded by the National Aeronautics and Space Administration.
This research has made use of the SIMBAD database,
  operated at CDS, Strasbourg, France.


\end{document}